\documentclass[12pt]{iopart}
\usepackage{iopams}
\usepackage{psfig}
\usepackage{graphicx}

\begin{document}

\title{Low noise cryogenic system for measurement of the
Casimir energy in rigid cavities}
\author{ Giuseppe Bimonte\P$\bigtriangleup$,  Detlef Born\S\P \footnote{Present address: NEST-CNR Scuola Normale Superiore Pisa, Piazza dei Cavalieri, 56126 Pisa, Italy}, Enrico Calloni\P$\bigtriangleup$,
Giampiero Esposito\P, Uwe Huebner\dag, Evgeni Il'ichev\dag,
Luigi Rosa\P$\bigtriangleup$,  Francesco Tafuri\S\P, Ruggero Vaglio\P$\bigtriangleup$ }

\address{\P INFN, Sezione di Napoli,
Complesso Universitario di Monte S. Angelo, Via Cinthia, Edificio 6, 80126
Napoli, Italy}
\address{$\bigtriangleup$ Dipartimento di Scienze Fisiche,
Complesso Universitario di Monte S. Angelo, Via Cinthia, Edificio 6, 80126
Napoli, Italy}
\address{\S Seconda Universit\`a di Napoli,via Roma 29, 81031 Aversa (CE), Italy}
\address{\ddag Institute for Photonic Technology,
Postfach 10 02 39 - 07702 Jena, Germany}

\begin{abstract}
We report on preliminary results on the measurements of
variations of the Casimir
energy in rigid cavities through its influence on the
superconducting transition
of in-cavity aluminium (Al) thin films.
After a description of the experimental apparatus we report on
a measurement in presence of thermal photons, discussing its
implications for the zero-point photons case. Finally, we show 
preliminary results for the zero-point case.
\end{abstract}



\section{Introduction}

In spite of great theoretical and experimental advances
\cite{Moste,Carugno}, the Casimir effect still faces important
unsolved questions. In Gravitation and Cosmology the problem of
reconciling the vacuum energy density and its interaction with the
gravitational field, known as the cosmological constant problem
\cite{Weinberg,Ishak}, remains unsolved. In flat space-time,
theoretical disagreements still remain on the dependence of the
Casimir energy (and force) on the geometrical shape of the bodies
and on the possibility that some geometries
exhibit positive energy and repulsive force \cite{Moste}. All 
experiments performed so far have been addressed to measure the
forces between (two) separate bodies, not directly the (variation
of) energy in a rigid body. Apart from the fundamental interest of
a direct measurement of the energy, the use of  rigid cavities
could make it easier to test different shapes. The interest is not
only theoretical: repulsive forces might build a bridge between
the Casimir effect and nanotechnology. On this basis, we recently
proposed the experiment Aladin2, to measure the variation of
Casimir energy in a rigid cavity, occurring when a thin metallic
film, constituting one of the plates of the cavity (the other
plate being made of a normal metal) becomes superconducting
\cite{Bimon1}. The experiment is made possible by the observation
that, under suitable conditions, the variation of vacuum energy in
the transition (due to the change in the film reflectivity)
$\Delta F^{{\rm cav}}$ can be comparable with the condensation
energy $E_{{\rm cond}}$ of the superconducting film, and therefore
it may sensibly affect the transition itself \cite{Bimon1,Bimon2}.
In particular, the parallel critical field $H_{\|}(T)$ required to
destroy the superconductivity of the film can be remarkably
different from the critical field of a bare film so that, to prove
this effect, the experiment measures the difference in the values
of $H_{\|}(T)$ for a film which is part of a Casimir cavity with
respect to an equal but bare film. The magnitude of the effect
depends also on the roughness of the plates and on their
conductivity, and hence, in principle, our method might be used to
probe the dependence of the Casimir effect on these factors, which
are currently being investigated experimentally. We remark that
our approach, being based on rigid cavities, could be easily
extended to measure the energy of different shapes, like for
example granular superconductors. Finally we notice that the
scheme of our experiment arose as a solution to the problem of
modulating the Casimir energy in a rigid cavity \cite{Callo1},
while introducing in the system the minimal amount of energy. This
goal is part of a more speculative project, aimed at verifying
whether or not Casimir energy gravitates, by weighing a Casimir
apparatus whose energy is modulated in time \cite{Callo1}.

\section{Expected signal}

The  Casimir cavity,  shown in Fig. 1, is composed of a thin Al
superconducting layer of thickness $D$, an intermediate dielectric
thin layer (Al$_2$O$_3$) of thickness $L$ and a third thick
metallic layer made of gold (Au) or silver (Ag). (See below for more details).
\begin{figure}
\centering
\includegraphics[width=0.48\textwidth]{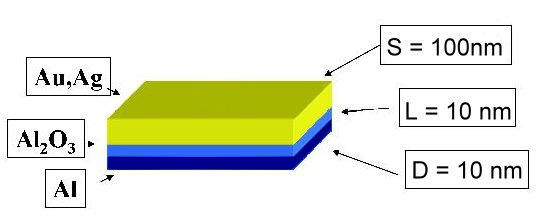}
\caption{Schematic of a Al-Al$_2$0$_3$-Au cavity.} \label{fig:Immagine6}
\end{figure}
\noindent  The expected effect is  shown in Fig. 2 where the shift
$\delta t = 1 -t$ of the reduced critical temperature $t =
T/T_{c0}$  is plotted against the critical field
$H_{\|}$; $T_{c0}$ is the zero-field transition
temperature. The upper curve  is for a bare superconducting thin
film while the lower curve  is for the Casimir cavity. For a bare
film with  $1-t\ll 1$, as it is our case, the critical field
$H_{\|}$ follows the well known equation \cite{Tinkam}:
\begin{equation}\label{eq:H}
    H_{\|}=
    H_{0}\sqrt{24}\frac{\lambda(0)}{D}\sqrt{1-t},
\end{equation}
\noindent which neglects the effect of nucleation, which applies to 
a very thin film.  Here, $\lambda$ is the penetration depth
and $H_{0}$ is the bulk zero temperature critical field. On
inverting Eq. (\ref{eq:H}), one obtains a parabolic dependence of
$\delta t$ on $H_{\|}$, corresponding to the upper curve of Fig.
2.
\begin{figure}[t]
    \centering
        \includegraphics[width=0.48\textwidth]{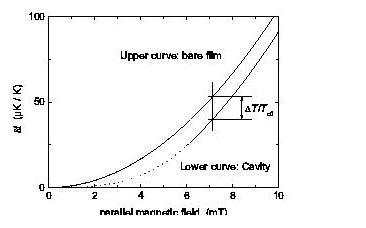}
        \caption{Simulation of the expected signal for a bare
thin Al film of thickness $D=14\,$nm (upper curve) and for a
cavity composed by a similar Al film, covered by a $6\,$nm
dielectric layer and a $100\,$nm Au mirror (lower curve).}
    \label{fig:Graph1}
\end{figure}
\noindent In a   cavity the dependence is expected to differ from
the bare film case in the region $\delta t<<1$, corresponding to
low applied fields. As shown in Fig. 2 the expected difference
$\Delta T$ in the transition temperature for low applied fields is
in the range of a few $\mu$K. (for higher values of the field,
not shown in the plot, the two curves approach each other)
\cite{Bimon1,Bimon2,Callo2}. In the region of very small applied
fields,  a deviation from the parabolic curve is expected by
continuity arguments (dashed line in Fig. 2); in this latter
region the change in free energy becomes comparable with the
condensation energy and no perturbative calculation can be done.
The choice of the material is crucial, and a compromise is usually
required to encompass  various needs. On the one hand, low
$T_c$ materials (low condensation energy) are preferable because
their transition is more perturbed by vacuum energy effects
\cite{Bimon2}. On the other hand, the longer cooling times
typically required  when working with (very) low $T_c$ materials
could  make it difficult to increase the statistics. In this first
experiment Al has been chosen has a good
compromise. Another important parameter is the plasma frequency
$\Omega_p$ of the normal plate; larger values of $\Omega_p$
imply a larger conductivity and ensure a better confinement of the
photons within the cavity. It was shown in Ref. \cite{Bimon2} that
a very good material could be Be ($\Omega^{Be}_{p} = 18.9$ eV),
but difficulties in obtaining cavities with this material led us
to use Au ($\Omega^{Au}_{p} = 9.0$ eV) or Ag ($\Omega^{Ag}_{p} =
8.2$ eV), resulting in a loss of approximatively  a factor of 2 in
the expected signal \cite{Bimon2}.  The values  of other
electrical parameters, such as the electrons mean free paths both
in the normal metal and in the superconductor, do not affect
significantly the expected signal \cite{Bimon2}. While the signal
does not depend on the area of the plates, the optimal thickness
$D$ of the superconducting metal turns out to be of a few
nanometers. Numerical calculations show that the variation of
vacuum energy in the transition has the following dependence on
the gap L:
\begin{equation}\label{free_energy}
  \Delta F_E^{(C)}(L) \propto\frac{1}{1+(L/L_0)^\alpha} ,
\end{equation}
\noindent where $\alpha = 1.15$ and $L_0 \simeq 10$ nm. As we see,
as soon as $L$ becomes smaller than $L_0$, $\Delta F_E^{(C)}(L)$
approaches a finite value and therefore, once $L$ is in the region
of few nanometers, no substantial gain can be obtained by using
smaller gaps. It can be shown that the signal  is practically
independent of the dielectric material, and  we chose the Al
native oxide Al$_2$O$_3$. In order to overcome spurious effects
due to different alignment of the cavity and the bare film with
respect to $H$, the cavity and the bare film are placed on the same
chip.

\section{Sample preparation, cryogenic apparatus and measurement method}

The films and cavities, made by IPHT (Institute for Photonic 
Technology - Jena - Germany), are obtained by depositing the Al on
the whole wafer (silicon, 3 inch diameter). The only difference between the
bare thin film and the cavity is the presence of a metallic layer
above the dielectric layer for the cavity. Finally the chip is cut
in samples that include, within a distance of  less than a mm,
several films and cavities.   Considering our parameters, as we reported elsewhere
\cite{Callo2,CalloBarca}, the applied magnetic field $H$ must be
very homogeneous, to within $10^{-4}$ over the distance $d$
between the cavity and the film ($d\approx 0.5\,$mm).  This
requirement has been fulfilled by a home-made coil system composed
by a  main coil and two lateral correction coils.  The alignement of
the field and sample is performed by  mounting rigidly the sample holder and the coil system.  If  $\theta$ is the angle between the magnetic field and the sample, following \cite{Tinkam}, for small angle $\theta$,  a perpendicular component of the field 
 adds a linear term in the dependence $\delta t(H)$, resulting approximately in
\begin{equation}
	\delta t \approx \frac{D^{2}}{24\lambda^{2}(0)H_{0}^{2}}H^{2} + \frac{\sin\theta}{H_0}H.
\end{equation}
Since the film and the cavity are deposited at close distance on the same chip, and since the magnetic field is very homogeneus, 
possible differences in the values of  $\theta$ and $H$ between the film and cavity can be neglected and therefore a non-vanishing $\theta$  just yields a common additive term
which does not affect the differential measurement.  
 
We used films with different
areas, ranging from about $20\times20\,\mu$m$^{2}$ to
$100\times100\,\mu$m$^{2}$, to verify that the effect does not
depend on the area. Inspection of   the samples with the atomic
force microscope showed that the actual thicknesses of the Al
layer     and the of the oxide gap are of   14 nm and 6 nm
respectively. The roughness is about 1 nm for both. Measurements
of different samples do not show any significant difference. The
cryogenic system is based on the HelioxVL $^3$He commercial cryostat from Oxford
Instruments, reaching a base temperature
of $250\,$mK, inserted in a home-made dewar equipped with magnetic
screening, to isolate the sample from undesired environmental
fields. The usual ferrite-filters at $300\,$K and 1K-pot levels
and  a cascade of two isolation copper-powder filters
\cite{Devoret} at the level of the 1K-pot and the $^3$He-pot
guarantee efficient EM screening.  
\begin{figure}[b]
    \centering
        \includegraphics[width=0.48\textwidth]{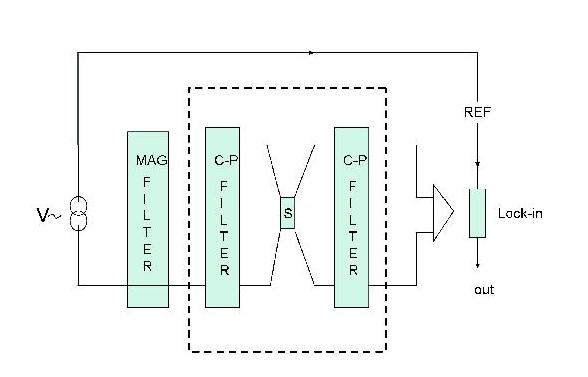}
        \caption{Schematic of the setup used for the measurement of sample resistance: the dashed box indicates the cryogenic environment,
        housing the sample and the copper powder filters for the four measurement lines.
        The driving electronic and the mixer, which are at room temperature, are also shown.}
    \label{fig:SchemaMisura}
\end{figure}
The measurement method for film and cavity resistances is a
standard four-wire measurement, as shown in Fig. 3. The temperature is
measured by a Cernox CX-1070-BR as part of a Wheatstone bridge. Both measurements 
are done with lock-in technique.  The actual measurement of $R(T)$ is
performed by fixing the external field, driving the temperature
from low to high and measuring $R(\tau)$ and $T(\tau)$ ($\tau$ is time) at the
sample during the transition. The cool down is always performed 
in zero magnetic field to avoid magnetic flux trapping.
Each single transition measurement
takes approximately 20 minutes. $R(T)$ is finally recovered to
evaluate the shift with respect to zero field case. The Al
resistivity ratio of the cavities and films analyzed in the
present work is $R(T=300\,$K$)/R_0(T=1.6\,$K$)\simeq 2$,
compatible with values quoted in the literature, corresponding to
a mean free path of about 15 nm \cite{Son}. The samples
resistances vary from $200\,\Omega$ to $350\,\Omega$. The
transition temperature at zero field, $T_{c0}$, is taken at the
maximum of the derivative of $R(T,H=0)$ and used as the
normalization factor in $\delta t$. In order to minimize the error
in the measurement, $\delta t (H_{\|})$
is obtained as the difference of temperature not  just at a given
point on the transitions with and without applied field (i.e. for
a given resistance $R$), but as the average of the differences for
all points of the transition (i.e. for all values of $R$):
\begin{equation}\label{eq:expData}
    \delta t=\frac{1}{T_{c0}}\cdot\overline{T(R,H_{\|}=0)-T(R,H_{\|})}.
\end{equation}
\noindent The average is  taken over the points with
$0.2 < \frac{R}{R_{N}} < 0.8$, where $R_{N}$ is the resistance
in the normal state.

\section{Measurement in the presence of thermal photons}

In order to test the cavity efficiency in removing photons
affecting the transition, we performed a first measurement in the
presence of room-temperature thermal photons. This has been done
by removing the cascade of copper powder filters and allowing the
external rf and microwave photons enter the vacuum chamber.
As pointed out in \cite{Bimon2}, the  important region of
frequencies $\nu_{{\rm eff}}$ contributing to  our effect is
around a few times $2 k_{B}T_{c}/h$. In the absence of a detailed
theory, as a first guess one can imagine that the presence of
extra thermal photons from the environment at temperature $T_E=300$K
gives rise to a temperature shift $(\Delta T)'$ that is larger
than the previous $\Delta T$  by a  factor $M$ equal to the ratio
between the black-body densities $E_{T_E}(\nu_{{\rm eff}})$ and
$E_{T_c}(\nu_{{\rm eff}})$ of electromagnetic energy at frequency
$\nu_{{\rm eff}}$, at temperatures $T_{E}$ and  $T_c$
respectively. Upon taking, to be definite, $h\nu_{{\rm eff}}= 10
k_{B} T_{c}$, we obtain
\begin{equation}
M=\frac{E_{T_E}(\nu_{{\rm eff}})}{E_{T_c}(\nu_{{\rm eff}})}
\approx   \frac{2}{{\rm e}^{\frac{10 T_c }{T_{E}}} - 1} \approx
\frac{T_E}{5\, T_c} \approx 40 .
\end{equation}
From this estimate we expect that this measurement (which, apart
from the filters, is  the same as the one described earlier),
should produce a signal similar to the one described in Fig. 2,
with $\Delta T$ increased by a factor  of some tens. As shown in
Fig. 4, this is what we indeed  found  in our measurements: the
curves relative to the bare film and the cavity, for low $H$, are
separated by a gap which reaches a maximum of about $(\Delta T)
'=300\,\mu$K. We point out that, although it must be considered as
a rough estimate, this measurement suggests a value of $\Delta T$
in the final experiment close to the most conservative expectation
of $\Delta T=10\,\mu$K (or even less).

\begin{figure}[h]
    \centering
        \includegraphics[width=0.48\textwidth]{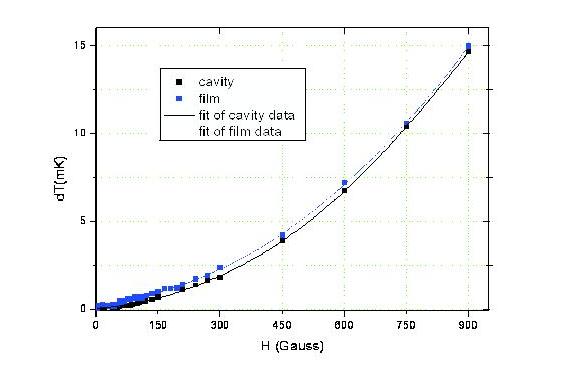}
        \caption{Results for the measurements without copper-powder filters
        (the lower curve is for the cavity, the upper for the bare film). The
        behaviour is qualitatively similar to the theoretical prediction in Fig.
        2,
       but the shift is enhanced from the presence of $T_E=300$K thermal photons:
        the maximum shift
        in temperature is about $300\mu$K, much higher than $1-10 \mu$K
        expected for zero point case. }
    \label{fig:CurveSenzaFiltro}
\end{figure}

\section{First results in the measurement of Casimir energy variations}

\begin{figure}[h]
   \centering
        \includegraphics[width=6.0cm]{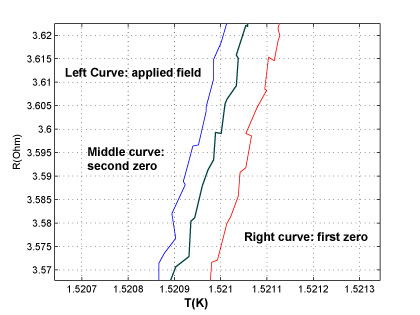}
    \caption{Example of a measurement triplet: a zoom of the resistence versus temperature for two transitions
in zero field and one with $\mu_0H=7.2\,$mT.
         The mean shift is $80\,\mu$K.}
    \label{fig:zoom}
\end{figure}

In the presence of copper powder filters, the major sources of
noise   is the electronic noise at the read-out amplifier. This
noise has a ``fast'' component (with time scale of one second)
that determines a statistical error   of a few $\mu$K, and a slow
thermal drift (linear in time, about $-50\,\mu$K per hour) that
produces a $\delta t$  proportional to the time elapsed between
two measurements.   To  correct for the thermal drift, at equal
time intervals before and after each measurement with applied
field, we perform two measurements in zero field, and then we take
for $\delta t$ the average of the two $\delta t$'s. We performed the measurement on three different samples,
without finding significant differences. An example of
such a measurement is shown in Fig. \ref{fig:zoom}: the
intermediate curve and the right curve were taken in zero field,
while the left curve is with an applied field of $7.2\,$mT. It
should be noted that the shifts in temperature for the three
curves  are only   a few tens $\mu$K. This procedure is repeated
for all applied fields.

\begin{figure}[h]
    \centering
        \includegraphics[width=0.48\textwidth]{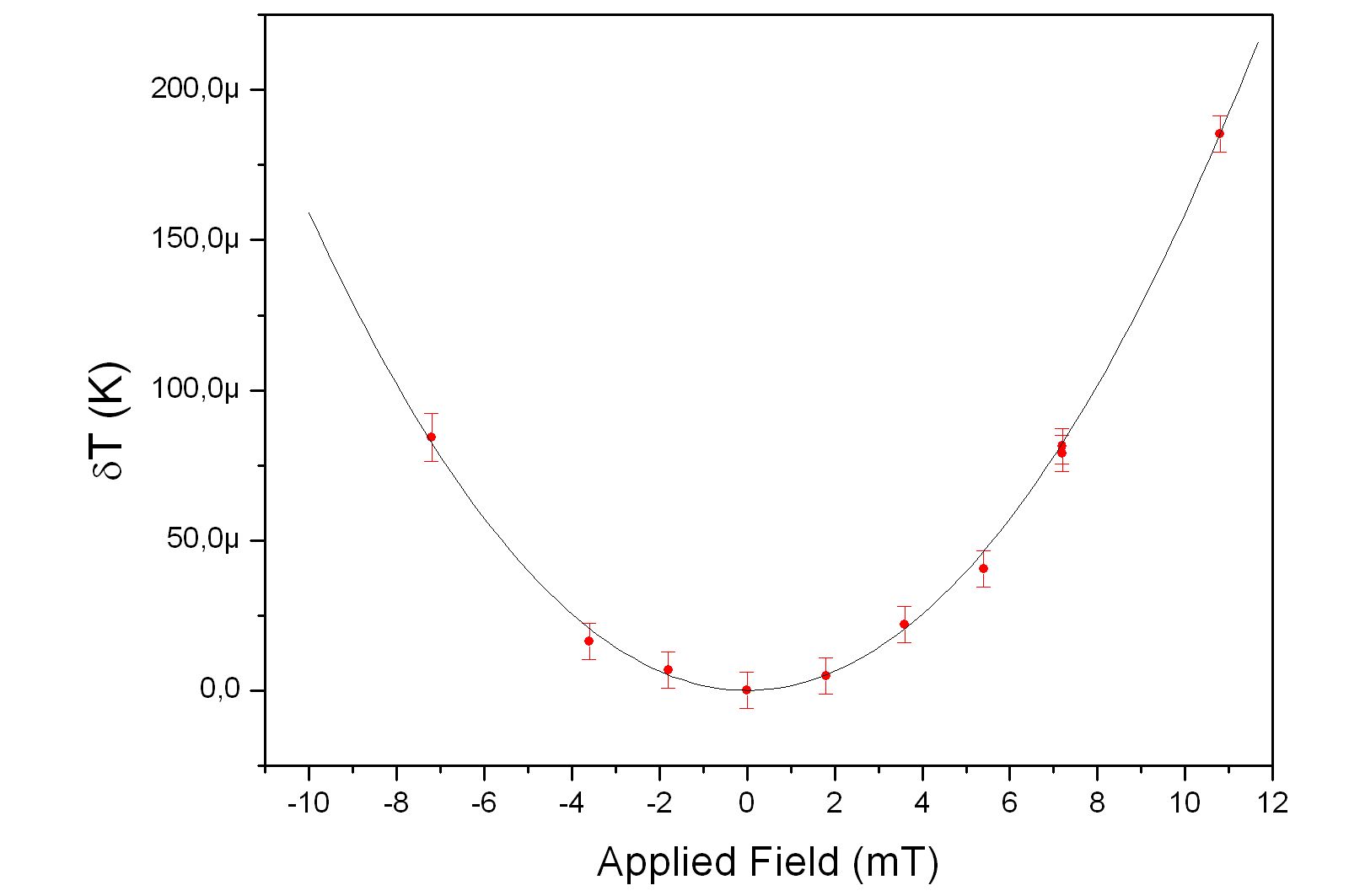}
        \caption{Experimental low-field points for a bare film. 
         The solid line is a parabola obtained from a fit of high-field measurements (data not shown).
         The error bars are of 6 $\mu$K, as estimated in the text.
         The good agreement indicates that there are no unexpected sources of noise }
    \label{fig:Parabola}
\end{figure}

The   data for a typical bare film are reported in Fig.
\ref{fig:Parabola}. The parallel field is applied in opposite
directions to  rule out  possible hysteresis.  We find that the
theoretical parabolic dependence of $\delta t$ on the applied
field is recovered for fields as low as a few mT, corresponding to
changes of critical temperature of about $10\mu$K. The
sensitivity has been roughtly estimated to be about $6\,\mu$K by comparing several measurements of $\delta t$ 
under the same conditions, performed at some days of distances.
\begin{figure}[h]
    \centering
        \includegraphics[width=0.48\textwidth]{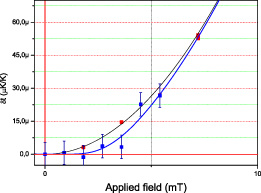}
        \caption{Results of the final measurements. The lower curve is the cavity, the upper is for the bare film.
        (Error  bars are $6\mu$K, see text.) The cavity shows a different behavior at very low fields (as expected from the theory).}
    \label{fig:results}
\end{figure}
It is important to stress that the solid line  was
obtained by fitting high-field measurements (data not shown),
which are less affected by the possible presence of EM noise or
field-film misalignment. The good agreement of the solid line with
the low-field measurements indicates that there are no unexpected
sources of noise in the low field region.

An example of a complete measurement, for one of our samples, is
reported in Fig. \ref{fig:results}. Other samples do not show
significant differences at the present level of sensitivity. The
upper curve is the same bare-film parabola of Fig.
\ref{fig:Parabola}, while the lower curve is a fit on the cavity
data.  To help reading the figure, the bare film data are reported
without errors. The cavity curve lies below the bare film curve,
and it shows a non-parabolic behavior, in qualitative agreement
with theoretical expectations. Moreover the maximum displacement
between the two curves, of about $7 \mu$K, is of the expected
magnitude. However, it is also comparable with the present
sensitivity and therefore the result of these measurements, while
encouraging, is far from being conclusive. In addition to this,
the two curves appear to merge much faster than expected
theoretically, and the reason for this is presently not clear.

In order to increase the sensitivity,  a new cryogenic apparatus
is being built, allowing to reach lower temperatures (about 20 mK)
and to use softer superconductors. We expect that the
signal-to-noise ratio will improve, as discussed in section 2, to
a sufficient extent to definitively confirm or disprove the
foreseen effect.

\end{document}